 \documentclass[final,5p,times,twocolumn,authoryear]{elsarticle}

\usepackage{amssymb}
\usepackage{subfig}
\usepackage{multirow, multicol}
\usepackage{xcolor}

\makeatletter
\def\ps@pprintTitle{%
  \let\@oddhead\@empty
  \let\@evenhead\@empty
  \let\@oddfoot\@empty
  \let\@evenfoot\@empty
}
\makeatother

\begin{document}

\begin{frontmatter}



\title{Negotiating Comfort: Simulating Personality-Driven LLM Agents in Shared Residential Social Networks}


\author[first]{Ann Nedime Nese Rende\corref{cor1}}
\ead{nedime.rende@bilkent.edu.tr}
\cortext[cor1]{Corresponding author: Ann Nedime Nese Rende}
\author[second]{Tolga Yilmaz}
\ead{tolga.yilmaz@cs.ox.ac.uk}
\author[first]{Özgür Ulusoy}
\ead{oulusoy@cs.bilkent.edu.tr}
\affiliation[first]{organization={Department of Computer Engineering, Bilkent University},
            postcode={06800}, 
            city={Ankara},
            country={Türkiye}}
\affiliation[second]{organization={Department of Computer Science, University of Oxford},
            postcode={OX1 3QD}, 
            city={Oxford},
            country={U.K.}}

\begin{abstract}

We use generative agents powered by large language models (LLMs)
 to simulate a social network in a shared residential building, driving the temperature decisions for a central heating system. Agents, divided into Family Members and Representatives, consider personal preferences, personal traits, connections, and weather conditions. Daily simulations involve family-level consensus followed by building-wide decisions among representatives. We tested three personality traits distributions (positive, mixed, and negative) and found that positive traits correlate with higher happiness and stronger friendships. Temperature preferences, assertiveness, and selflessness have a significant impact on happiness and decisions. This work demonstrates how LLM-driven agents can help simulate nuanced human behavior where complex real-life human simulations are difficult to set.

\end{abstract}




\end{frontmatter}




\section{Introduction}
\label{intro}

Social network simulations are widely utilized to model the interactions between people, often relying on agent-based modeling to represent the relationship between people and their environment. In these simulations, the actions of the agents are selected from a predefined set of rules specified by the modeler. While this rule-based approach allows for a clear definition of the decision process and provides control over the outcomes, it also introduces a limitation. The predefined rules may not be able to model various dimensions of human behavior, such as irrational decision-making, and restrict the knowledge of agents to what is encoded by the modeler.  

Large language models (LLMs) are trained on a vast amount of data, mostly obtained from web pages \citep{llmBasedSurvey}. Learning from this human-generated data allows the models to have a level of real-world knowledge and reduces the amount of external information that is required to be given to perform various tasks. Moreover, with their abilities such as reasoning and role-playing, LLMs have previously been shown to have significant capabilities of simulating human-like behavior. Generative agents, as introduced in \cite{genAgents}, rely on LLMs to generate agent behaviors, based on agent-specific memory about the agent's identity, interactions with the other agents, and the environment. These agents reflect on their previous experiences, plan their next actions, and perform those actions. In addition to their ability to provide context-sensitive responses, generative agents can form relationships, coordinate with each other, and spread information through dialogues among themselves. 

Building upon the capabilities shown in this work, a growing number of studies have been integrating LLMs into agent-based simulations across diverse fields, including public health, social sciences, and robotics. To exemplify, GABM Epidemic \citep{gabmEpidemic} utilizes generative agents to model the spread of COVID-19. AgentSims \citep{agentSims} models a virtual town and provides an environment for the evaluation of LLM agents. AgentSpeak \citep{agentSpeak} simulates the electric vehicle adoption decisions depending on personal information and government incentives. \cite{roboticsConsensus} apply LLMs for a multi-robot collaboration task where agents' aim is to find a middle point to gather. 

In the field of Building Energy Modeling (BEM), LLMs have previously been employed to generate input files \citep{eplusLLM} for the EnergyPlus simulation software \citep{energyPlus} as well as to produce daily human activity data using generative agents \citep{dailyActivitiesLLM, knowledgeDistillation, privateLLMEnergyCons}. In addition, social network-based approaches have been utilized to inspect the dissemination of energy-saving behaviors \citep{greenConsumptionBehavior, passiveAndActiveInteractions}. The effect of personality traits and preferences on building energy consumption has also been explored in prior studies \citep{thermostatPreferences, liu2021}. However, these works do not consider the integration of social network structures with generative agents for a personality-driven simulation of decision making in the context of BEM. Our approach introduces a novel perspective on modeling residents in households and aims to represent human decisions in a more comprehensive manner. 

We choose building energy as the context for social decision-making due to its multifaceted nature. Energy consumption is influenced by a mix of external factors (e.g., weather conditions), social ties, personal traits, and preferences. A simulation of decision-making within the building energy domain allows the inspection of the relationship between these factors over time. On each day of our simulation, agents determine the temperature for the central heating system through both family-level and building-level polls. In the family-level decision stage, we provide the agents with the information of the heater preference of family members, their personality traits, the current happiness level ranging from 1 to 100, and the weather outside. In this context, agents are tasked with deciding on a temperature for the heater and updating their happiness levels  
depending on the final decision on the previous day. The personalized prompts are sent to the LLM, and answers from this stage are included in the building-level decision prompts. 

We label the agents depending on their roles as a Family Member or a Family Representative. The family representatives are the final decision-makers who present their votes to the building poll. In this stage, agents are also provided with closeness levels to friends and temperature preferences of those friends. With this updated context, agents are asked to provide a final degree decision, and are given the option to update their closeness levels with friends. 

With this simulation scenario, we execute simulations in three different settings with varying distributions of positive personality traits. These settings are labeled as all positive, all negative, and 50\% positive. Through regression analyses at the network level, we examine the impact of this distribution on all our network-level variables, such as average happiness level, cost, and average degree of friendships. At the node level, we determine the most significant factors for the individual temperature choices and happiness levels through panel data regression. 

We implement, execute, and visualize our simulations using the Crowd tool \citep{crowd} from our previous work, which is a social network simulation framework that allows quick simulation setup either as a Python library or through its graphical user interface. Crowd provides features such as interactive graph and chart visualizations, as well as automated data collection. In this paper, we also explain how we employ Crowd for generative agent-based social network simulations. 

Our contributions can be summarized as follows:
\begin{itemize}
    \item We present a methodology for modeling energy-related decision-making in shared residential settings by integrating generative agents into social network simulations.
    \item Employing three sets of experiments, we inspect the effects of varying distributions of positive and negative personality traits on agents' happiness levels and evolving friendships.
    \item Through network-level and node-level statistical methods, we analyze the relationship between social network structure, personality traits, and heater preferences on the temperature suggestions and happiness levels of agents.
    \item  We provide an example use of the Crowd framework for generative agent simulations in a social context.
\end{itemize}

In the following sections, we first provide a brief summary of similar studies and our social network simulator, Crowd (Section 2). Then, we explain the methodology employed in our study to create families, the simulation scenario, and the simulation logic for agent decisions (Section 3). We then present the results of our simulations and regression analyses (Section 4). Following the results, we discuss the outcomes of our simulations, drawing out the limitations and further considerations (Section 5). Finally, we conclude with a summary of the study's key aspects and a discussion of future work (Section 6).

\section{Related Work}

\subsection{Social Networks, Personality and Preferences in Building Energy Modeling}
In the field of building energy modeling (BEM), factors like the presence of the occupant in the house, interactions with household systems (e.g., heating or lighting), and personal information such as energy awareness and purchasing preferences are considered to be highly influential in the analysis of energy consumption \citep{occupantBehaviorSurvey}. For this reason, agent-based modeling and simulation (ABMS) is commonly employed to model the occupants and the building environment, as well as their interactions. While some studies focus on agents' interaction with the environment, others inspect the impact of social factors on agents' consumption behavior. In such cases, embedding agents within a network environment enables researchers to take advantage of social network analysis methodologies to study the diffusion of positive behavior \citep{greenConsumptionBehavior, passiveAndActiveInteractions}, 
or the impact of regulation policies such as government subsidies for the adoption of new technologies to promote carbon-friendly consumption \citep{greenConsumptionBehavior}.

Moreover, various works analyze the effects of personality on energy-saving behavior through the data collected from surveys \citep{liu2021, hotelGuests}. The residents are divided into groups of varying characteristics with respect to their personality trait scores, and these groups illustrate different patterns of energy-saving behavior. In the residential house setting,  \cite{liu2021} found that residents of the positive group obtain higher energy-saving behavior scores compared to people in other categories, while introverted people choose to stay passive. In addition to the groups of different personalities,   \cite{hotelGuests} further inspect the impact of household habits on energy-saving behavior in a hotel setting. 

Another categorization of residents is made based on their thermostat preferences to simulate the influence of these preferences on overriding automated temperature selections \citep{thermostatPreferences}. In this study, agents are classified as either ``average" or ``tolerant" based on their behavioral patterns. 

In our study, we employ social networks to model both family and friendship dynamics, with a focus on reaching an agreement within agents, rather than inspecting the diffusion of energy-saving behavior. To simplify the task of understanding personality traits by the LLM, we represent these traits using descriptive adjectives corresponding to positive and negative scores, rather than using numerical values. Furthermore, we extend the heater preferences of agents to include five levels (cold, cool, neutral, warm, and hot), which adds more variety to agents' personal profiles. 

\subsection{Use of LLMs in Building Energy Modeling}

Based on the concepts introduced in the paper that proposed generative agents \citep{genAgents}, various studies in diverse research disciplines have utilized LLMs to simulate the daily lives of agents. In the field of BEM, obtaining daily activity data through simulated agents enables researchers to have access to large amounts of data without raising privacy concerns. Through modifications to the simulation engine provided in \cite{genAgents}, \cite{privateLLMEnergyCons} generate daily activities of agents in households. By filtering the activities that consume energy from the agent activities, daily energy consumption is calculated. \cite{dailyActivitiesLLM} utilize generative agents in a smart home simulation context, and explore various fine-tuning methods to improve the performance of smaller models on the daily activity generation task. To model the real world with more diversity and accuracy, \cite{knowledgeDistillation} further extend this task by incorporating various types of family structures, energy usage profiles, and weather conditions, conducting experiments based on six different countries. In their study, LLMs are utilized to generate the families, their daily activities, the electricity consumption of each activity, and the weather conditions. 

These studies form the daily schedule of agents by generating hourly actions and considering interactions with different appliances. In our study, we only focus on the heating system, with each simulation step representing a single day. To inspect the social dynamics in energy-related decisions, we model a central heating system that maintains a constant temperature during the day, determined by the requests of residents in the building. In addition to the agent-specific factors, such as age, personality, and heater preferences, the temperature selection of agents is further influenced by the suggestions from family members and friends. 

\subsection{Crowd}
Crowd \citep{crowd} is a social network simulation framework that allows quick simulation setup through YAML configurations, Python code, and a graphical user interface (GUI). 
Crowd provides four types of network classes tailored to varying simulation tasks. For our study, we select the Custom Simulation Network environment where the methods the user passes in are executed at the specified times. The simulation graph and all other data are saved to the file system automatically for future visualization or statistical analysis. Additionally, Crowd's GUI facilitates interactive network visualizations, data aggregation, and chart generation based on the collected simulation data. 

In this study, we utilize these facilities of Crowd to run all our simulations, merge data from different simulations, and visualize our results with line and bar charts by simply using the selectors in its GUI.  

\section{Methodology}
\subsection{Creating Families}
\label{creating_families}

\begin{figure}[!t]
    \centering
    \includegraphics[width=0.95\linewidth]{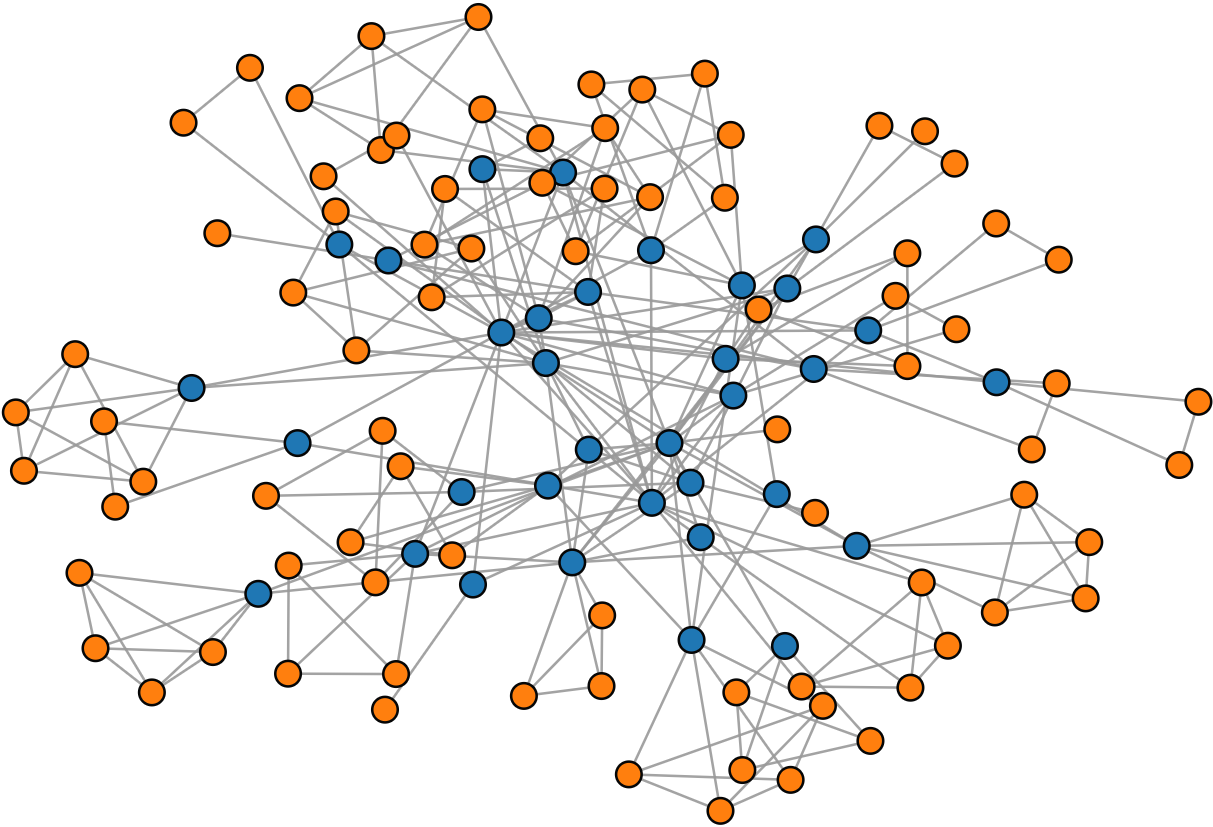}
    \caption{Network visualization of created families where nodes are colored according to the roles (blue: Family Representative, orange: Family Member).}
    \label{fig:network_roles}
\end{figure}

\begin{figure}[!t]
    \centering
    \includegraphics[width=0.95\linewidth]{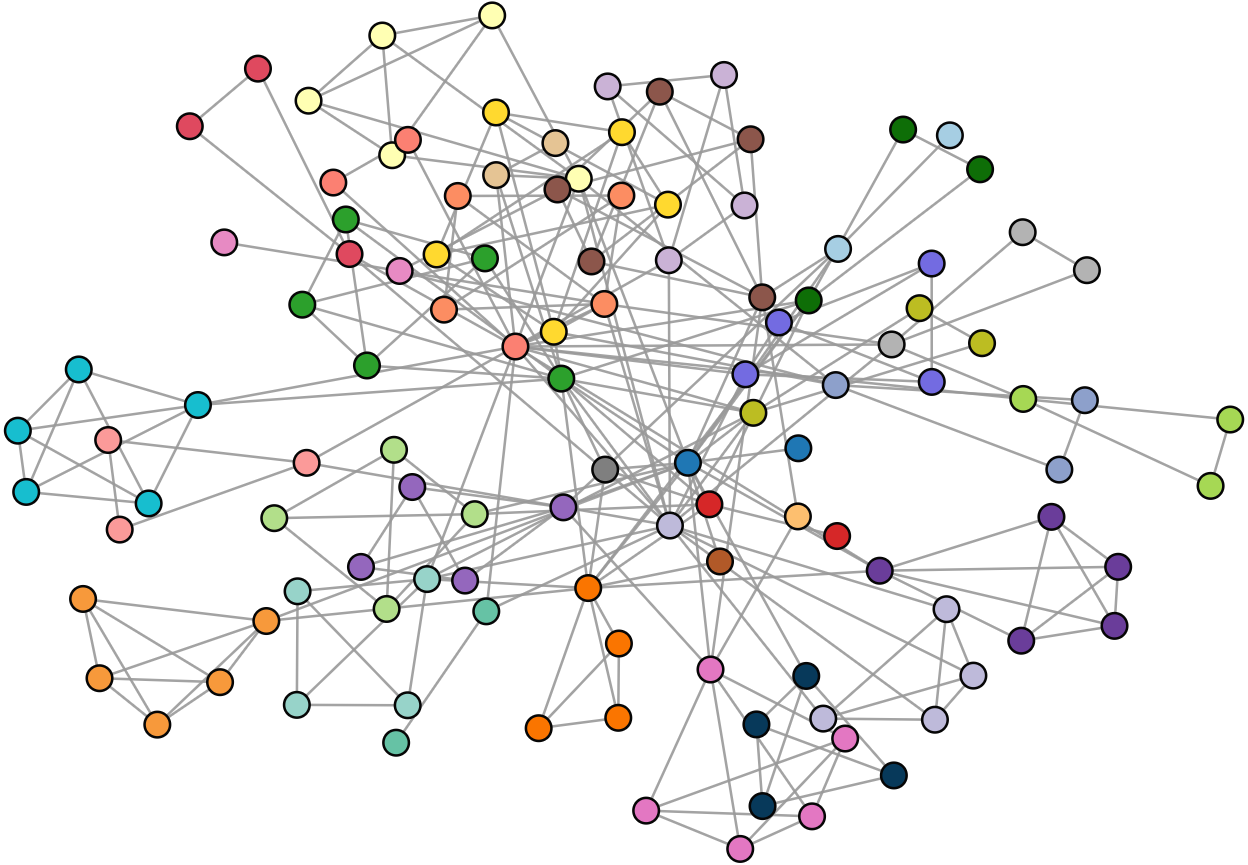}
    \caption{Network visualization of created families where nodes are colored according to their family IDs.}
    \label{fig:network_families}
\end{figure}

\begin{table*}[!ht]
    \centering
    \caption{Node parameters related to personality and associated adjectives.}
    \begin{tabular}{r c c c } 
    \hline
        \textbf{Parameter Name} & \textbf{Big 5 Trait (Facet)} & \textbf{Positive Adj.} & \textbf{Negative Adj.}\\ \hline
        Angry Hostility & Neuroticism (N2)& Easygoing & Easily-angered \\ \hline
        Impulsiveness  &  Neuroticism (N5) & Self-controlled & Impulsive\\ \hline
        Assertiveness & Extraversion (E3) & Assertive & Passive\\ \hline
        Feelings & Openness to Experience (O3) & Emotional & Unemotional\\ \hline
        Alturism & Agreeableness  (A3) & Selfless & Selfish \\ \hline
        Compliance  & Agreeableness (A4) & Cooperative & Uncooperative\\ \hline
        Deliberation  & Conscientiousness (C6) & Cautious & Careless\\ \hline
        Environmentalism & x & Environmentalist & Not environmentalist\\ \hline
        Frugality & x & Frugal & Wasteful\\ \hline
    \end{tabular}
    \label{tab:personality_traits}
\end{table*}
We first load Zachary's Karate Club network \citep{karateClub}, which is an undirected real-world network widely utilized in social networks research. It consists of 34 nodes and 78 edges. We categorize these nodes as the ``Family Representatives", who will be presenting the family's daily temperature preference in the building poll. The links between Family Representatives represent the friendship relationship. The weight of these edges is utilized in the simulation as ``closeness level".

For each Family Representative, we randomly assign $n$ family members, where $n$ is selected from the range $[0,4]$. All family members are connected to each other, creating a fully connected subgraph represented by the Family Representative's ID. The new network consists of 116 nodes and 246 edges. Fig. \ref{fig:network_roles} illustrates the created network according to the roles of the nodes (Family Representative or Family Member), and Fig. \ref{fig:network_families} shows the family relationships within in the network, where the fully connected family sub-graphs are observable through the coloring of the nodes.

Following this step, each node in the extended graph is assigned a name, age, and a set of personality traits. These traits are selected from the Big 5 Traits' Facets \citep{big5Facets}, based on their relevance to the simulation's scope. Facets such as \textit{Angry Hostility} and \textit{Feelings} are included to observe their influence on agents' happiness levels, while \textit{Impulsiveness, Assertiveness, Alturism, Compliance,} and \textit{Deliberation} are expected to affect the temperature choices and closeness levels between friends. Instead of assigning numerical values to represent traits, we use a pair of opposing adjectives—one positive and one negative—for each facet. This approach simplifies the interpretation of the facets for the LLM. The assignment ratio of these facets to nodes is varied for the different simulation settings.

In addition to the node parameters representing the personality types, agents are also labeled based on their care for the environment, spending habits, and heater preferences. A list of node parameters related to personality and the associated adjective pairs is given in Table \ref{tab:personality_traits}. The node and edge parameters generated at this stage are used to personalize the prompts sent to the LLM, which will act as a person with these traits and simulate human behavior with a broader range of possibilities compared to traditional rule-based methods.

\subsection{Extracting Weather Data}
To observe the behavior of the generative agents from a broader perspective as much as possible within 30 days, we select a month of the year with fluctuating temperatures. This corresponds to a period between February 15 to March 16 from the Typical Meteorological Year (TMY) data obtained utilizing \textit{pvlib} \citep{pvlib}. We select Ankara, the capital of Türkiye, as the location for our experiments, which can be extended to other cities in future studies. 

Instead of using the hourly weather data provided, we take the average temperature for each day. While this approach does not consider the day and night temperature differences, it significantly reduces the execution time by limiting the simulation to one iteration per day rather than 24. Therefore, we can utilize the data of a longer period (30 days) and run our simulations with different settings, while each simulation takes notably less time to complete. We execute our simulations on an L4 GPU through Google Colab, and a simulation of 30 days takes approximately 13 hours. 

\subsection{Simulation Scenario}
\label{sim_scenario}

Our simulation is based on a building with 34 apartments and a central heating and cooling system. Each day, the families come together to decide on a temperature setting for the system. For this decision, they consider their own temperature preferences, their family members' preferences, and the degree outside. Each family member presents a degree, affected by these properties as well as their personality traits. The average of all the members' choices is taken and given to the Family Representative. If a person lives alone, they only consider their own preferences for this step. 

Each person is also assigned a happiness value. This value is set to 100 at the beginning of the simulation, and each day, agents are asked about their level of happiness with respect to their previous day's choice and the final temperature set in the building. 

After the family's decision, the next step for a Family Representative is to pass another degree choice to the building poll. In this stage, each Family Representative is also informed about their friends: Their closeness level, friend's family's degree choice for that day, and friend's degree suggestions for the last three days. The Family Representative makes a second decision based on their friends' and family members' choices, their own temperature preference, and personality. Each day, these agents may decide to increase or decrease the closeness values with their friends, updating the weight of the edge between them. 

\subsection{Simulation methodology with Crowd}
\label{sim_method_with_crowd}
For our simulations, we choose Crowd's Custom Simulation Network as the environment. This network type requires the simulation logic to be defined by the modelers in Python methods, which are then passed to Crowd for execution. Crowd provides four intervals for the execution of custom methods: \textit{before-iteration} (e.g., iteration setup methods), \textit{every-iteration-agent} (called for each agent in random order), \textit{after-iteration }(e.g., data collection methods) and \textit{after-simulation} (used to finalize the simulation or collect end results). 

\begin{figure}[!t]
    \centering
    \includegraphics[width=0.95\linewidth]{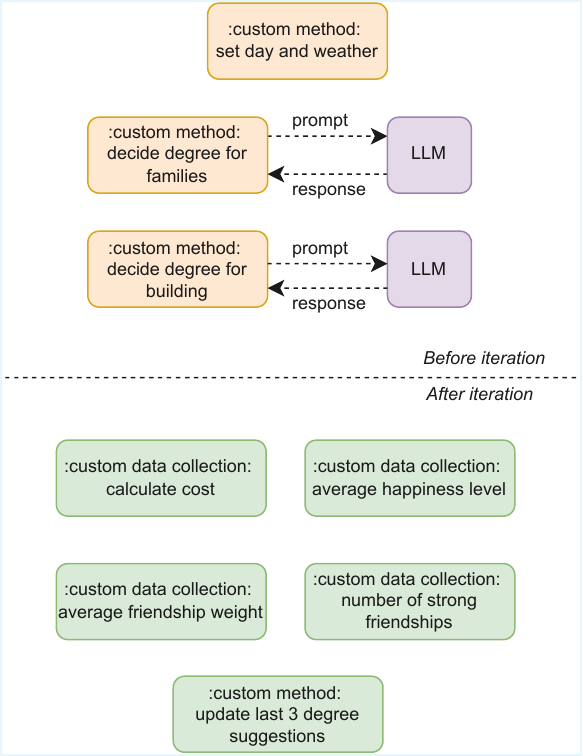}
    \caption{The methods passed to Crowd framework and their execution order.}
    \label{fig:crowd_methods}
\end{figure}

Our custom methods for this study, and their execution periods are provided in Fig. \ref{fig:crowd_methods}. Instead of using \textit{every-iteration-agent} method call interval of Crowd, we directly loop over the agents with respect to their family IDs or perform model-level calculations directly within these functions. 

Environment variables in Crowd are named as ``network-parameters". In our simulations, we store the current day and its temperature as network-level parameters, updating them daily. After this initialization, we have two phases of temperature decision-making, following the simulation scenario explained in the previous subsection. In Phase 1, we ask the LLM to act as the agents in the families and provide its choices. In Phase 2, LLM makes decisions for the Family Representatives, and we set the building's temperature as the average of their degree suggestions.

Each day, we calculate the cost using Eq. \ref{cost-calculation} given below:

\begin{equation}
\label{cost-calculation}
    cost = | d_{set} - d_{outside} |\times C
\end{equation}

In this equation, $C$ is a constant that represents the cost of increasing the building temperature by one degree. As the aim of this study is to present a simulation methodology rather than provide precise estimations of real-world energy consumption, we set $C=1$ for simplicity. However, future works may investigate the impact of fluctuating energy unit prices per degree $C$ on consumption behavior.

The data for average friendship weight and the number of strong friendships are also collected at the end of each iteration. For this calculation, only the nodes with the ``friend" connection (i.e., Family Representatives) are considered. The edge weights between friends are initially obtained from the Karate Club network, and they may change with respect to agents' decisions in each iteration. When the weight of an edge between two friends is larger than 3, it is considered as a strong friendship. 

Lastly, before moving on to the next iteration, we call the \textit{update last 3 choices} function, which updates the associated node parameter that is given to the LLM in the second phase prompt.

\subsection{Agent decisions with LLMs}
In our simulations, we assign the role of decision-making for each agent to an LLM, instead of following a traditional rule-based model. We choose Mistral 7B \citep{mistralLLM} and apply 8-bit quantization, aiming to complete the queries to LLM more quickly while still maintaining the use of an open-source model. 

\begin{figure}[!t]
    \centering
    \includegraphics[width=0.68\linewidth]{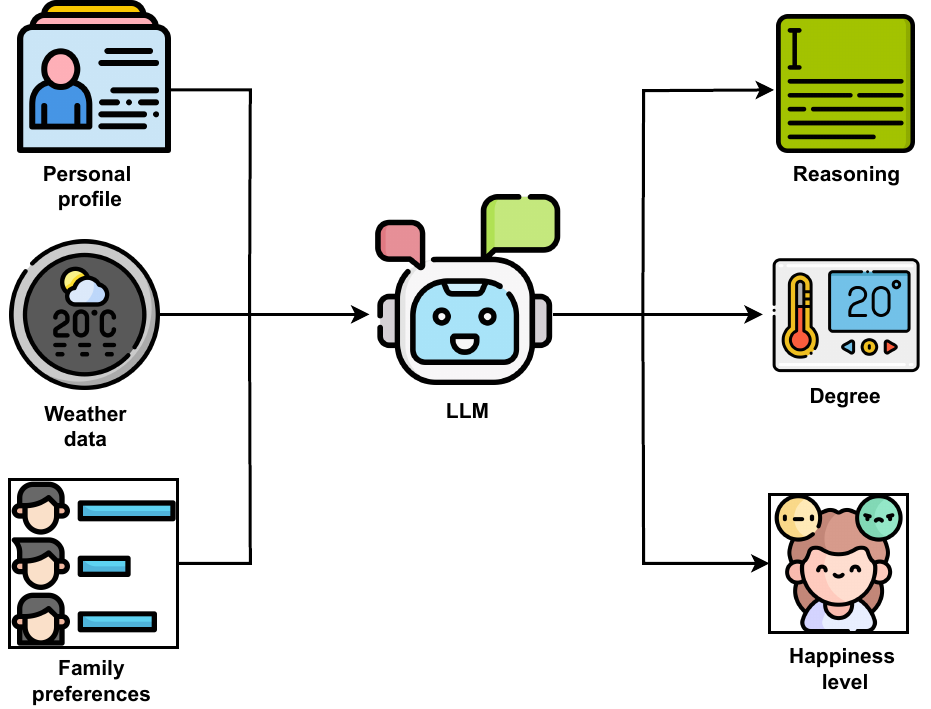}
    \caption{Inputs provided to LLM and outputs obtained in the first phase.}
    \label{fig:sim-phase-1}
\end{figure}

\begin{figure}[!h]
    \centering
    \includegraphics[width=0.68\linewidth]{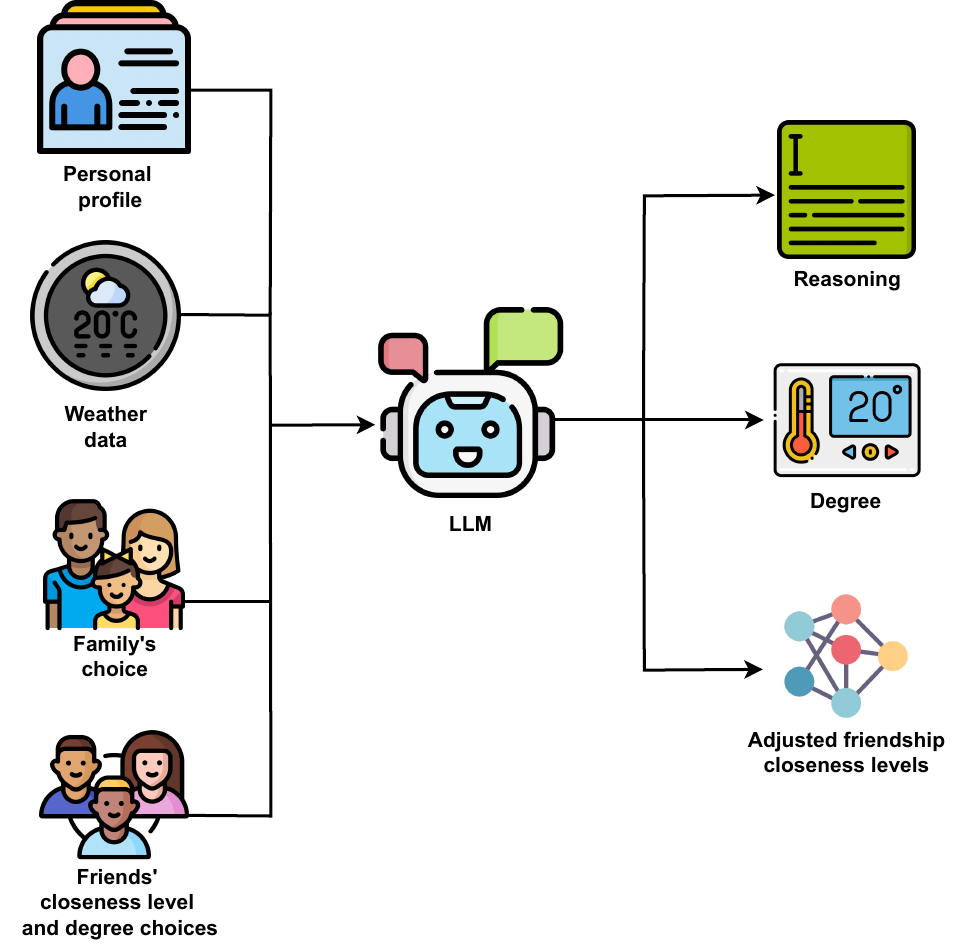}
    \caption{Inputs provided to LLM and outputs obtained in the second phase.}
    \label{fig:sim-phase-2}
\end{figure}

\begin{figure*}[!h]
    \centering
    \includegraphics[width=0.99\linewidth]{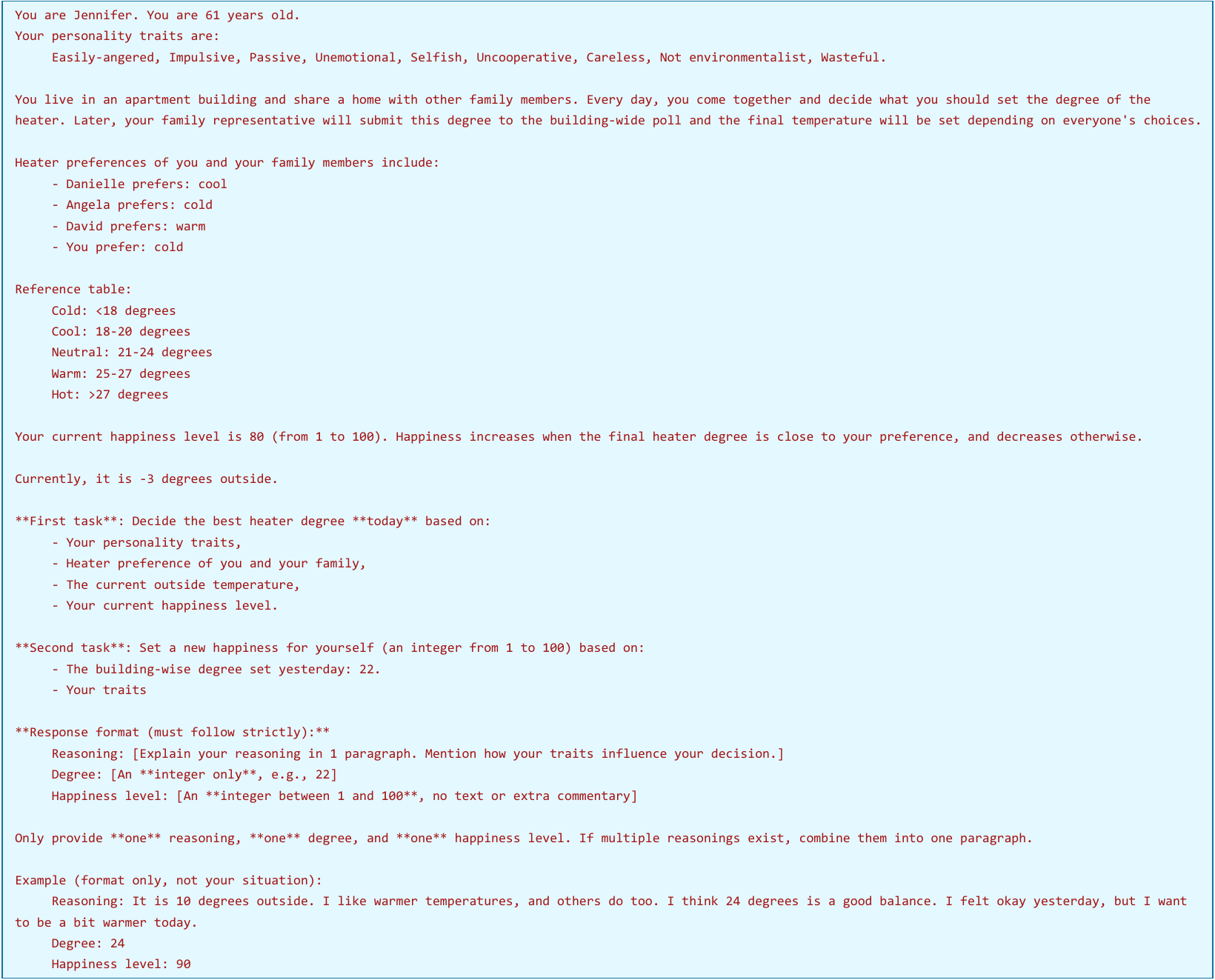}
    \caption{Phase 1 sample prompt.}
    \label{fig:phase1-prompt}
\end{figure*}

\begin{figure*}[!ht]
    \centering
    \includegraphics[width=0.99\linewidth]{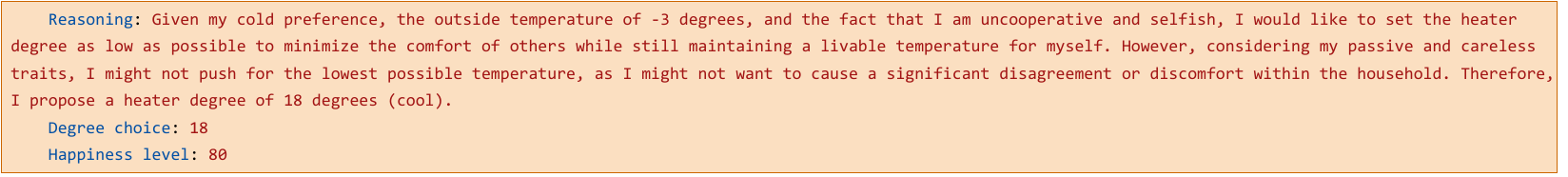}
    \caption{Phase 1 sample output.}
    \label{fig:phase1-output}
\end{figure*}

\begin{figure*}[]
    \centering
    \includegraphics[width=0.97\linewidth]{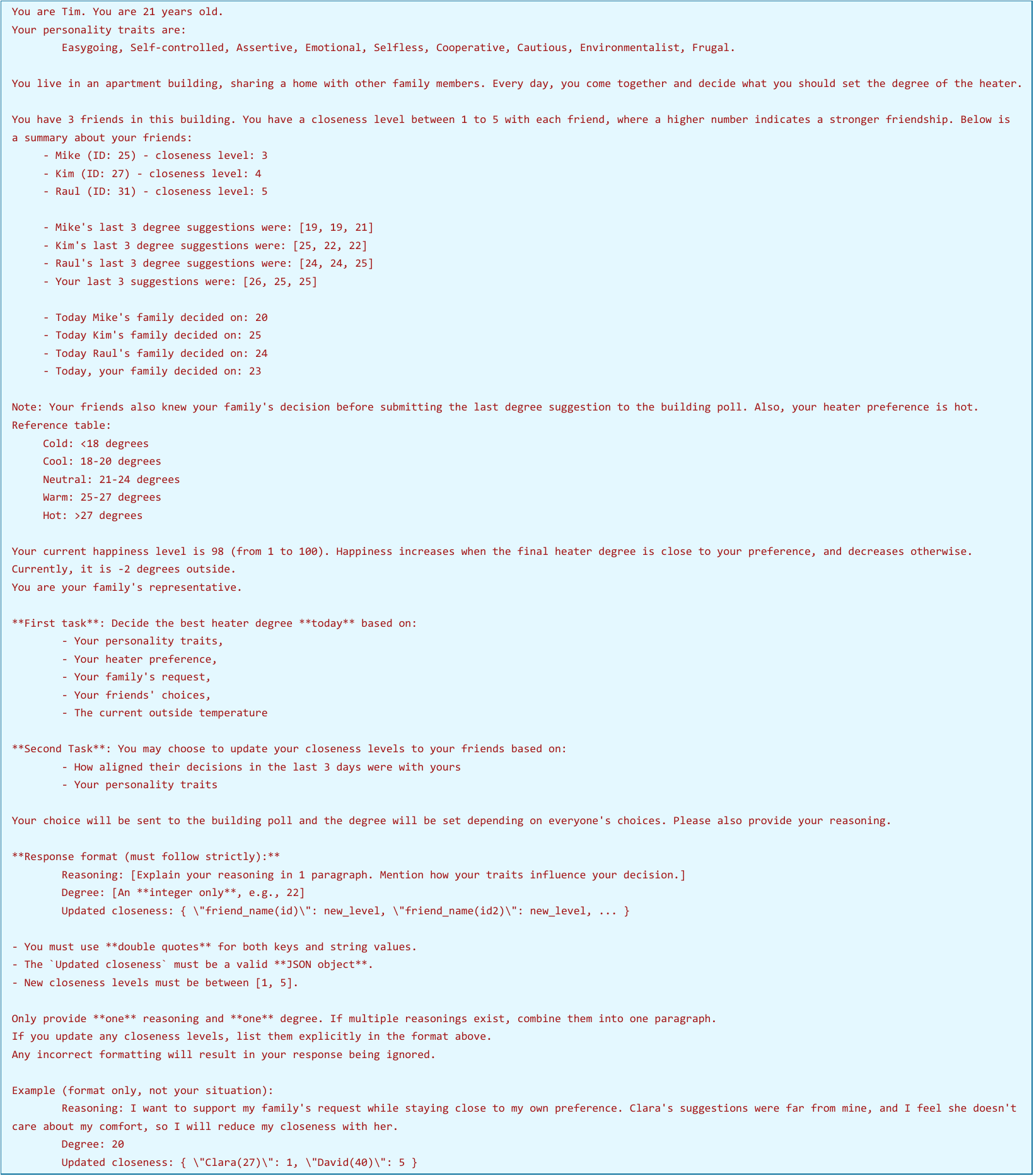}
    \caption{Phase 2 sample prompt.}
    \label{fig:phase2-prompt}
\end{figure*}

\begin{figure*}[]
    \centering
    \includegraphics[width=0.97\linewidth]{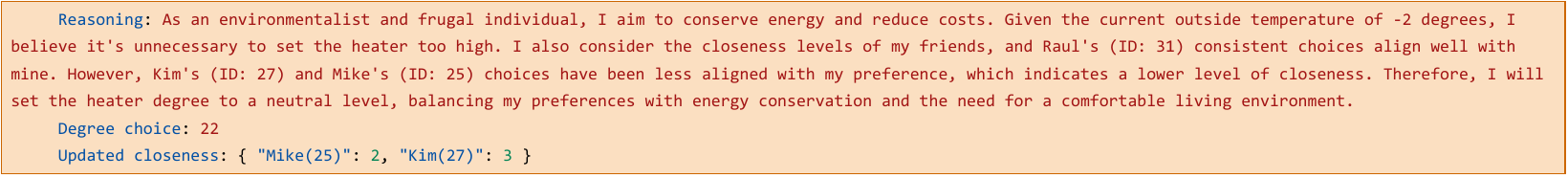}
    \caption{Phase 2 sample output.}
    \label{fig:phase2-output}
\end{figure*}

Illustrations of the first and second phases of the simulation are provided in Figures \ref{fig:sim-phase-1} and \ref{fig:sim-phase-2}, respectively.
In addition, an example prompt for the first phase is shown in Fig. \ref{fig:phase1-prompt}. In this stage, agents are provided a list of family members' heater preferences, their current happiness levels, and the weather information. Moreover, a reference table for the heater preferences categories and the corresponding degree ranges is provided. In our initial experiments, we observe that without this reference, LLM may suggest nonviable degrees such as 1 or 100 as the output. To further enhance our prompt, we separate the tasks, itemize the decision factors, and add an example output within the instructions. These suggestions, provided by ChatGPT \citep{chatgpt}, improve the quality of the reasoning provided in each query's output and show that larger models can be utilized to polish the input for smaller models for these types of tasks. 

\begin{figure*}[!ht]
    \centering
    \subfloat[]{
        \includegraphics[width=0.45\linewidth]{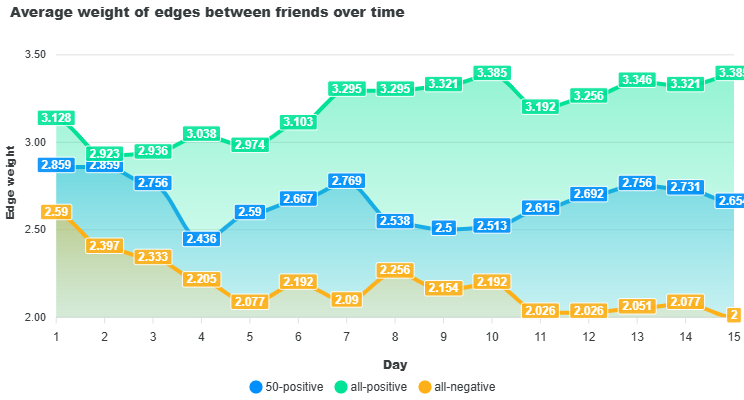}
        \label{fig:edge_weights}
    }
    \hspace{0.02\linewidth}
    \subfloat[]{
        \includegraphics[width=0.45\linewidth]{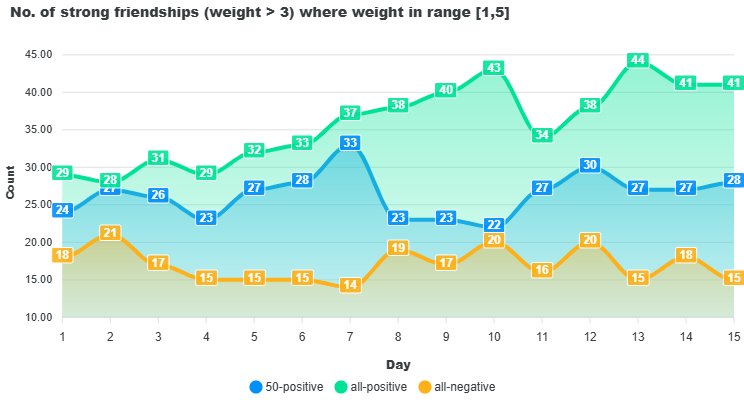}
        \label{fig:strong_friendships}
    }
    \\
    \subfloat[]{
        \includegraphics[width=0.45\linewidth]{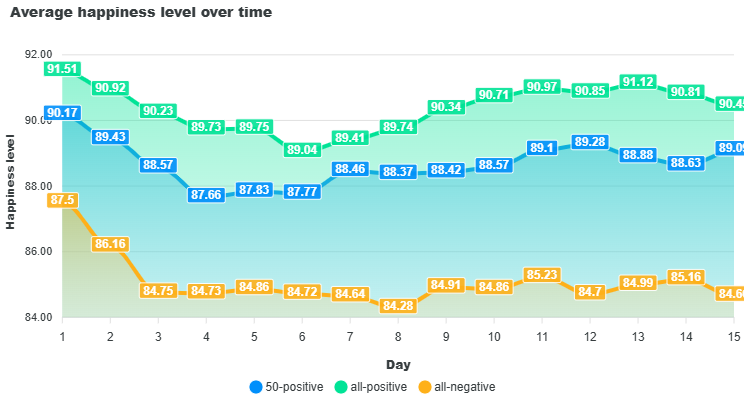}
        \label{fig:happiness_levels}
    }
    \hspace{0.02\linewidth}
    \subfloat[]{
        \includegraphics[width=0.45\linewidth]{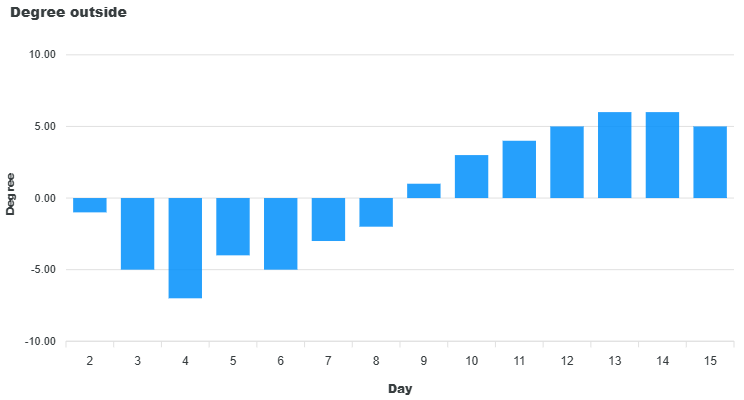}
        \label{fig:degree_outside_15_days}
    }
    \\
    \subfloat[]{
        \includegraphics[width=0.45\linewidth]{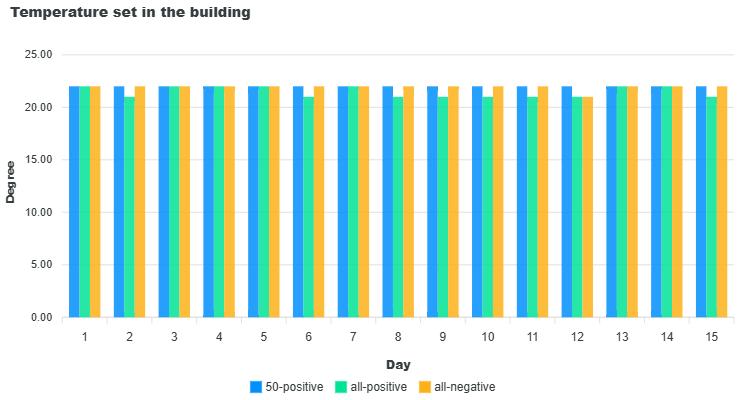}
        \label{fig:temperature_set}
    }
    \hspace{0.02\linewidth}
      \subfloat[]{
        \includegraphics[width=0.45\linewidth]{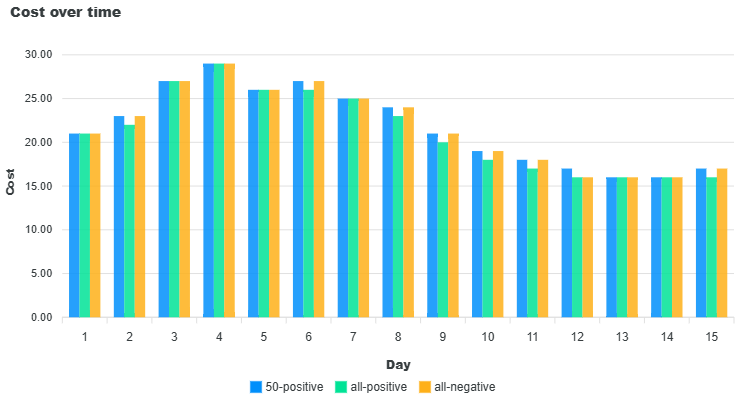}
        \label{fig:cost}
    }
    \label{fig:network-charts}
    \caption{Charts generated with Crowd. (a) Average weight of edges between friends over time. (b) Number of strong friendships (weight $>$ 3 where weight in range $[1,5]$). (c) Average happiness level over time. (d) Degree outside. (e) The temperature set in the building. (f) Cost over time.}
\end{figure*}

The output provided by the LLM for the sample prompt is given in Fig. \ref{fig:phase1-output}. As instructed, the LLM returns a reasoning for its decision based on the personality traits of the agent and the preferences of the agent's family. We extract the degree choice and happiness levels of each agent, and as explained in Section \ref{sim_scenario}, provide each family's average suggestions to the Family Representative in the second phase.  Through our experiments, we observe that the order and structure of the friends' information in the prompt given in Fig. \ref{fig:phase2-prompt} impact the responses. When all information related to a friend is given together without categorization by content, the LLM may evaluate any one of the categories while ignoring others. 
Hence, we categorize the information with respect to its contents, and specifically place the agent's suggestions and preferences immediately afterward for better comparison. Since the number of closeness updates is not predetermined, we explicitly instruct the LLM to return a JSON object for the ease of parsing. 

The output corresponding to the example prompt is given in Fig. \ref{fig:phase2-output}. In this sample, it can be seen that the agent decides to decrease the closeness levels with the friends of differing choices, while keeping the highest closeness level with the friend suggesting similar temperatures. Moreover, despite the hot heater preference of the agent and the higher temperature choices of family and close friends Kim and Raul, the agent ultimately suggests a temperature falling into the lower neutral category to conserve energy and reduce costs because of his personality traits.

\section{Results}

\begin{table*}[!ht]
\centering
\caption{OLS slope coefficients (with standard errors) over 15 days.}
\label{tab:network-slopes}
\begin{tabular}{ r l l l }
\hline
\textbf{Metric} & \textbf{All Positive} & \textbf{50\% Positive} & \textbf{All Negative} \\
\hline

Avg. friendship weight   & 0.0298 (0.0059)*** & -0.0145 (0.0088) & -0.0295 (0.0059)*** \\  \hline
No. of strong friendships       & 1.0786 (0.2021)*** & 0.0214 (0.1648)  & -0.0464 (0.1381)  \\  \hline
Cost                     & -0.8071 (0.1770)*** & -0.7750 (0.1729)*** & -0.7893 (0.1804)*** \\  \hline
Temperature set          & -0.0321 (0.0308) & 0.0000 (0.0000) & -0.0143 (0.0155) \\
\hline
\end{tabular}

\vspace{1mm}
\footnotesize{* p $<$ 0.05, ** p $<$ 0.01, *** p $<$ 0.001}

\end{table*}

To investigate the impact of personality traits on happiness levels, degree selections, and closeness levels between friends over time, we run our experiments in three different setups. While the initial network structure remains constant in all settings,  the main difference lies in the distribution of personality traits.  As explained earlier in Section \ref{creating_families}, we utilize adjectives derived from the Big 5 Traits' Facets and group them as either positive or negative. The derived adjectives are grouped together with their opposites, and assigned to agents in each simulation based on the selected \textit{positive trait percentage}. In our three setups, we select this\textit{ positive trait percentage} as 100\%, 50\%, and 0\%, respectively, and collect both network-level and node-level data to observe the effects of this distribution. While these experiment setups are limited and may not fully reflect the complexity of real-world societies, they provide an efficient way to observe the contrast considering the cost of running generative agent simulations with large networks and long iterations. 

\subsection{Network-level analysis}

At the network level, we examine how the structural features of the network, temperature-related factors, and average happiness levels of agents evolve over time. For this analysis, we first generate charts with Crowd, given in Fig. 
10. Then, to complement the visual representation of our results, we apply the Ordinary Least Squares (OLS) method to compute the regression slopes for each metric given in these charts. The slope values presented in Table \ref{tab:network-slopes} quantify the rate of change and highlight the differences across our three types of personality trait distributions. While the charts illustrate the daily fluctuations occurring with respect to the changes in outer factors such as the weather conditions (degree outside), the numbers in Table \ref{tab:network-slopes} provide a broad perspective on the overall trends. The p-values in this table represent how well the regressed line is able to represent the general trend of the changes of the corresponding variable. 

Cohesion in networks refers to the existence and strength of ties, or the level of connectedness among nodes, from a structural perspective \citep{networkCohesion}. Since the average friendship weight and the number of strong friendships are the metrics that rely on tie strength, they can serve as indicators of network cohesion.  Table \ref{tab:network-slopes}, Figures \ref{fig:edge_weights} and \ref{fig:strong_friendships} show that both metrics increase over time in the all-positive case, while they decline in the all-negative case. This suggests that these two opposing personality type cases can result in notably distinct dynamics of network cohesion. 

As explained in Section \ref{sim_method_with_crowd}, we calculate the cost as the absolute difference between the temperature set and the degree outside. Hence, the calculated cost is highly dependent on the weather outside, which can be observed from Figures \ref{fig:degree_outside_15_days} and \ref{fig:cost}. With the increase in the outside temperature, the cost decreases across all cases. 

Fig. \ref{fig:temperature_set} illustrates that, regardless of individual agent traits or preferences, the building temperature (determined by the average choice of the Family Representatives) consistently remains around 21–22 degrees across all simulations. This temperature falls under the Neutral range given as input to the LLM. This indicates that despite the agents do not communicate directly with each other to negotiate a temperature that is best for everyone, the collective decision still falls under a facilitative degree. While the building temperature is always set to 22°C in the 50\% positive case, we observe that both all-positive and all-negative agents occasionally vote for 21°C, which explains the differing coefficients shown in Table \ref{tab:network-slopes}. In those days, the decrease in cost is slightly impacted by this metric as well. 

Lastly, Fig. \ref{fig:happiness_levels} displays the changes in the average happiness levels over time. The mean happiness over the 15 iterations is 91.01 for the all-positive case, 89.33 for the 50\%-positive case, and 85.99 for the all-negative case. These values represent increases of 1.88\% and 5.84\% in the all-positive case compared to the 50\%-positive and all-negative cases, respectively. While the difference is more apparent in comparison to the all-negative case, it is relatively small when compared to the 50\%-positive case.

For another network-level analysis, we inspect the influence of the other variables on the average happiness level of agents, using OLS. For the regression, we combine the results obtained in the three settings and include the data of all 30 days collected for the 50\% positive traits case, while each of the other cases contributes 15 observations, summing up to 60 observations. 

\begin{table}[!t]
    \centering
    \caption{OLS Regression results to analyze the influence of network-level variables on average happiness level.}
         \resizebox{0.5\textwidth}{!}{%
         \begin{tabular}{ r c c }
        \hline
          \textbf{Parameter} & \textbf{Coefficient (Std. Error)} & \textbf{$p$-value} \\ \hline
          Average friendship weight &  4.723 (0.314)     & $ <0.001 $      \\ \hline
          Degree outside & 0.076 (0.023)       & $ <0.001 $      \\ \hline
        \end{tabular}
    }
    \label{tab:network-level}
\end{table}

Although we collect multiple network-related and temperature-related information in the simulation, we only utilize one of each category in this analysis to avoid correlation between the regressors. Hence, in Table \ref{tab:network-level}, we present the best combination of regressors: average friendship weight and degree outside.  Our results indicate that a 1-unit increase in the average friendship weight corresponds to a 4.72-point increase in happiness, while a 1-degree rise in external temperature (degree outside) is associated with a 0.076-point increase in happiness. 

In our study, the agents are prompted to keep their relationships closer with the friends whose choices align with theirs. Moreover, depending on their personality traits, agents may pay more attention to friends' choices with higher closeness levels and try to accommodate between their own preference, family's request, and friends' choices. This attempt at cooperation and higher levels of connectedness results in a higher level of average happiness level.

\subsection{Node-level analysis}

Since our aim for the node-level analysis is to inspect the effects of different personality traits and node-level SNA metrics on individual happiness level and degree choices, we exclude the 100\% and 0\% positive settings for this task. With the remaining data, we use a Correlated Random Effects (CRE) \citep{creModel} model to inspect both within-agent and between-agent variation in 30 days, for both happiness and degree choice analyses. We chose the CRE model over the standard random effects model as the latter was rejected by the Hausman test. However, the fixed effects model eliminates the time-invariant predictors, which are the main focus of our analysis.

During our simulations, we collect two time-variant types of continuous data: individual happiness levels and daily degree choices. On the other hand, the time-invariant categorical data include node parameters, such as personality types (Big 5 Facets, environmentalism, frugality), non-weighted centrality metrics, and heater temperature preferences. While estimating the impact of the other variables on the happiness levels, we include another regressor required for the CRE model, labeled as \textit{degree choice mean}. This is calculated individually for each agent by taking the average of their own degree choices, hence staying constant throughout 30 days. Similarly, we include the \textit{happiness level mean} while analyzing the effects on the degree choice. 

\begin{table}[!t]
    \centering
    \caption{Correlated random effects model to estimate happiness level on 30-day node level data.}
    \resizebox{0.5\textwidth}{!}{%
    \begin{tabular}{ r c c }
    \hline
      \textbf{Variable} & \textbf{Estimate (Std. Error)} & \textbf{\textit{p}-value} \\ \hline
      Degree choice &  0.2468 (0.0601)     & $<0.001$***   \\ \hline
      Degree choice mean & 3.8119 (0.0920)       & $<0.001$***      \\ \hline
      Assertive (E3) &  3.2413 (0.9494)     & $<0.001$***       \\ \hline
      Emotional (O3) &  1.0160 (0.9354)     & 0.2775      \\ \hline
      Selfless (A3) &  -2.6886 (1.2999)     & 0.0387*     \\ \hline
      Environmentalism &  2.0929 (1.2058)     & $0.0827^\dagger$   \\ \hline
      Warm temperature pref. &  -7.7128 (1.1907)     & $<0.001$***     \\ \hline
      Hot temperature pref. &  -13.584 (1.2503)     & $<0.001$***     \\ \hline
    \end{tabular}
    }
    
    \footnotesize{$\dagger$ p $<$ 0.1, * p $<$ 0.05, ** p $<$ 0.01, *** p $<$ 0.001}
    \label{tab:node-level}    
\end{table}

Table \ref{tab:node-level} shows the results of the regression on individual happiness levels with the best-performing estimators. With 116 nodes and 30 iterations, we have 3480 observations for this analysis. The results indicate that the most significant regressors are degree choice, agents' mean degree choice, assertiveness, and temperature preferences. 

The degree choice (the temperature suggestion of the agent for the day) and the agents' mean degree choice positively impact agents' happiness levels, indicating that people who suggest higher temperatures tend to achieve higher happiness. On the other hand, warm and hot temperature preferences have a negative relationship with the happiness level. While this may seem contradictory on the surface, the coefficients of these estimators and the reference information given to the LLM in the prompt for the temperatures provide a different point of view. 

As can be seen from the example prompts given in Figures \ref{fig:phase1-prompt} and \ref{fig:phase2-prompt}, warm and hot preferences of nodes indicate a preference for temperatures above 25°C. However, Fig. \ref{fig:temperature_set} shows that the temperature finally set in the building is either 21°C or 22°C in all iterations. These values fall under the neutral range, and as the gap between the agents' preferences and the neutral range increases, their dissatisfaction with this selection becomes more significant. The results from Table \ref{tab:node-level} illustrate that the agents having a warm temperature preference have happiness levels that are 7.7128 points lower, while those with a hot temperature preference experience a decrease of 13.584 points in happiness. This suggests that the positive relationship between the degree choice and happiness levels ($\beta = 0.2468$) holds true when the degree choice is not far from the neutral range or is still in the colder ranges. 

In terms of personality traits, assertiveness and selflessness have the highest significance. While assertiveness is positively associated with happiness ($\beta = 3.2413$), selflessness has a negative impact ($\beta = -2.6886$). As assertive agents aim to align the outcome with their own preferences, they tend to suggest temperatures that match their ideal range, which keeps their happiness levels high. On the contrary, selfless agents may make decisions far from their choices, considering the preferences of their family and friends more than their own, resulting in lower happiness values. 

Being an environmentalist also affects the happiness positively ($\beta = 2.0929$), however it is not as significant as the other personal traits. Although ranking as the third most important personality facet, being emotional is shown to have less significance. 

\begin{table*}[t]
    \centering
    \caption{Correlated random effects models to estimate degree choice on 30-day node-level data.}
    \begin{tabular}{ r c c c c }
    \hline
    \multirow{2}{*}{\textbf{Variable}} & \multicolumn{2}{c }{\textbf{Model 1}} & \multicolumn{2}{c }{\textbf{Model 2}} \\ \cline{2-5}
    & \textbf{Estimate (Std. Error)} & \textbf{$p$-value} & \textbf{Estimate (Std. Error)} & \textbf{$p$-value} \\ \hline
    Happiness level & 0.0202 (0.0049) & $<0.001$*** & 0.0202 (0.0049) & $<0.001$*** \\ \hline
    Happiness level mean & 0.2157 (0.0058) & $<0.001$*** & 0.2273 (0.0056) & $<0.001$*** \\ \hline
    Assertive (E3) & -0.5343 (0.2287) & 0.0196* & -0.4616 (0.2682) & $0.0854^\dagger$ \\ \hline
    Selfless (A3) & 0.5556 (0.3073) & $0.0707^\dagger$ & 0.5422 (0.3596) & 0.1317 \\ \hline
    Degree centrality & 3.7636 (4.2801) & 0.3793 & -- & -- \\ \hline
    Warm temperature pref. & 2.1054 (0.2745) & $<0.001$*** & -- & -- \\ \hline
    Cool temperature pref. & -- & -- & -1.0937 (0.3323) & 0.001*** \\ \hline
    Hot temperature pref. & 3.4629 (0.2871) & $<0.001$*** & 2.4978 (0.3282) & $<0.001$*** \\ \hline
    \end{tabular}\\
    \footnotesize{$\dagger$ p $<$ 0.1, * p $<$ 0.05, ** p $<$ 0.01, *** p $=<$ 0.001 }
    \label{tab:combined-node-level-degree}
\end{table*}

Table \ref{tab:combined-node-level-degree} shows the combined results for the CRE model on agents' degree choices with two models, differing in the regressors included. Consistent with the estimates from Table  \ref{tab:node-level}, happiness level and degree choice are positively associated. On the other hand, the sign of the coefficient $\beta$ changes for the personality traits assertiveness and selflessness. Assertiveness negatively impacts the degree choice ($\beta = -0.5343$), while being selfless increases the degree choice by 0.5556. This may be explained by the fact that assertive people tend to suggest lower temperatures to align with their own preferences, despite the needs and preferences of the other people in the building. In contrast, selfless people aim to keep the temperature in the neutral range for the comfort of their family and friends.

A key observation from the results presented in Table \ref{tab:combined-node-level-degree} is regarding the degree centrality. Degree centrality is a metric commonly employed in SNA to define a node's importance with respect to its number of connections. In our study, the nodes have both family and friend ties, and our calculation for this metric is inclusive of both types of connections. Although the centrality metrics or the full network structure is never given to the LLM in any phase, the nodes with higher degrees appear in more people's prompts as friends. Hence, these nodes are expected to influence more agents' decisions, making degree centrality a  significant regressor. However, with a p-value of  0.3793, our results indicate that the degree centrality of the nodes is not as influential as the other factors. 

\begin{figure}
    \centering
    \includegraphics[width=0.99\linewidth]{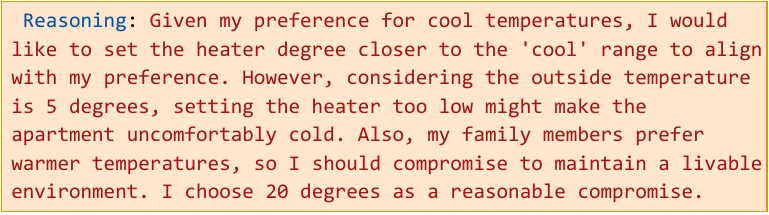}
        \caption{Example output showing the agents choosing the higher limit of their preference category considering their needs and other people's comfort.}
    \label{fig:phase1_compromise}
\end{figure}
Similar to their impact on the happiness levels, the temperature preferences remain significant for the \textit{degree choice}. Despite warm and cool temperature preferences have the same distance to the neutral category, their coefficients differ notably. While this may be a consequence of using different regressors in the model, we also observe this trend through the agents' responses. As can be seen in Fig. \ref{fig:degree_outside_15_days}, the temperature outside changes between -7 and 6. In the responses from the LLMs, it is evident that the agents acknowledge the need for heating and consider other people's needs. Hence, despite their cold or cool preferences, the agents may suggest a degree closer to the neutral range, going with the higher limit of their preference category. An example of this situation is given in Fig. \ref{fig:phase1_compromise}. A similar trend can be followed for the hot preference in Model 1. Moreover, the positive sign of the coefficient for warmer degree preferences and the negative sign of the coefficient for the cool preference align with the logical expectation.

\section{Discussion}
The impact of the Big 5 personality traits on network structure has been shown to be significant in various studies \citep{selden2018review}. \textit{Extraversion} is mostly associated with more connections (i.e., larger ego-networks), and \textit{agreeableness} is related to the maintenance of ties over time. Conscientiousness is shown to be significant for maintaining family ties and the positions of people in the workplace networks. People tend to connect to individuals with similar \textit{openness} levels, and \textit{neuroticism} is not strongly associated with network size and composition.

In our study, personality traits were distributed randomly without considering the positions of the individuals within the network. For a more realistic distribution, the degree of nodes can be considered for the trait assignments, or the simulation can be conducted on another real-life social network that includes personality labels for each node.

Our results indicate that assertiveness and selflessness are significant factors in the happiness levels in this scenario. As shown in Table \ref{tab:personality_traits}, these factors belong to the extraversion and agreeableness categories of the Big 5 personality traits. A related study conducts an analysis on the impact of personality on social support and subjective well-being among first-year students in a university \citep{zhuHappiness}. In contrast to our analysis on simulated data, their work is based on real-life data collected through surveys. They inspect the influence of personality traits on well-being through both direct and indirect paths. Through the directed paths, neuroticism shows a significant relationship with subjective well-being, while the impact of other traits (e.g., extraversion and agreeableness) are mediated through the intermediary network variables on the path (e.g., network size). 

To represent the Big 5 personality traits, \cite{zhuHappiness} use numerical scores. Although this is a common approach, it may not be the best method for the studies employing generative agents. Providing textual information instead of numbers may be more comprehensible to LLMs when evaluating the instructions to role-play and make decisions as a person with the given characteristics. At this stage, careful selection of the information included in the instructions is also essential to avoid possible misinterpretation. 
As explained earlier in Section \ref{creating_families}, we select personality trait facets in our simulation based on their relevance to the scenario. For instance, we only utilize ``Angry Hostility" and ``Impulsiveness" facets of the neuroticism trait. Despite neuroticism was found to significantly influence happiness levels in a prior work \citep{zhuHappiness}, the facets of neuroticism were not statistically significant in influencing the happiness level decisions made by the generative agents employed in our simulations. The positive association of extraversion with well-being aligns with our findings. However, we observe that agreeable individuals have lower happiness levels in this scenario, possibly as a consequence of prioritizing other people's comfort over their own preferences.

Regarding the network-level analysis of the relationship between the average friendship weight and happiness levels shown in Table \ref{tab:network-level}, our results are similar to those of \cite{zhuHappiness}, where closeness is positively associated with perceived social support, which is a highly significant estimator of subjective well-being.

In this study, we employed the Mistral 7B model to generate agent decisions for its efficiency and ease of use. Using larger and stronger models is expected to affect the decisions made by the agents, and hence the overall results. Future work would include doing a comparative analysis of utilizing different models and investigating how the social interactions take place with respect to them.

In the LLM-based generation of human daily activities and occupancy data for building energy modeling, the studies focus on hourly actions of the agents \citep{privateLLMEnergyCons, knowledgeDistillation}. The interactions of agents and the energy-consuming appliances are also calculated and visualized on an hourly basis. For instance, the heating and cooling systems are only turned on when there is at least a single family member at home. This approach, combined with the inclusion of the average cost of using certain appliances, allows for more realistic simulations. As our focus in this study is to show the impact of personality traits, preferences, and social network structure on the temperature decisions, we simplify the simulations by making each iteration represent a day, and by setting the cost of increasing the temperature of the building by a degree to a value of 1. By simply changing the value of the constant $C$ in Eq. \ref{cost-calculation}, researchers can achieve more realistic results for the given scenario. 

In addition, expanding the scope of the simulation to include more iterations, locations with varying climates, and attributes of family structure could lead to valuable insights. The importance of family structures and culturally sensitive activity generation is explored in \cite{knowledgeDistillation}, which can also be extended to our simulation scenario. Our current simulation model also assumes no communication or friendship between family members from different households, which limits the realistic representation of a community. 

As previously mentioned in Section \ref{intro}, \cite{genAgents} show that simulation with generative agents also exhibits emergent behavior such as information diffusion. Therefore, various studies focusing on the diffusion of energy-saving behavior over social networks and the effects of public interventions on such cases (e.g., \cite{greenConsumptionBehavior, passiveAndActiveInteractions}) can be extended with generative agent-based modeling.  

\section{Conclusion}

In this paper, we presented a simulation methodology that integrates generative agents with social networks to simulate personality-driven social dynamics in the context of building energy usage. By replacing traditional rule-based agent decisions with LLM queries, we take advantage of LLM's real-world knowledge and reasoning capabilities to generate more context-aware and human-like behaviors. In our prompts to the LLM, we provide the agent's personality traits, preferences, happiness levels, and their family members' preferences, and task the agent with selecting a temperature to set the heater for the day. Moreover, in the second stage of our simulation, we give the agents the closeness levels with friends and their friends’ previous temperature choices, and allow the agents to modify their closeness levels. Through three sets of simulation settings where we vary the percentage of positive personality traits assigned to the agents, we examine the effects of personality traits on happiness levels, average closeness levels between friends, and temperature choices. We utilize Ordinary Least Squares and Correlated Random Effects regression models to provide statistical analyses at both the network and node levels. 

Our results at the network level illustrate that the average happiness level of agents is positively influenced by both the average friendship weights and the weather outside. On the other hand, at the node level, we observe that the degree choice of agents is positively correlated with their individual happiness levels, as long as the suggested degree does not exceed the neutral range. With the final temperature set in the building being either 21°C or 22°C in all simulation settings, agents with non-neutral temperature preferences exhibit lower happiness levels, indicating dissatisfaction with the collective decision. Being assertive and environmentalist increases happiness, while being selfless results in a decrease. 

In our analysis of node-level degree choices, we observe that assertive agents suggest lower temperatures, while selfless agents propose higher temperatures. As expected, warm and hot temperature preferences are positively associated with the degree choices, while cool temperature preference has a negative correlation. 

As future work, we aim to extend the simulations across more iterations to observe the continuation of the behavioral trends, and to collect and analyze additional social network metrics for further analysis. In addition, we plan to experiment with more advanced LLMs with larger parameter sizes to inspect potential improvements in the reasoning of decisions, LLMs' ability to act as a person with the given traits, and understanding of the numeric information provided. A comparison of the social interactions across different LLMs could also provide a reference for future studies. Lastly, we plan to extend our simulation settings to include cities of varying climates and cooling systems in the summer season.






\bibliographystyle{elsarticle-harv} 
\bibliography{example}






\end{document}